\documentclass[11pt]{article}
\usepackage[preprint]{neurips_2020}
\pdfoutput=1
\usepackage{times}
\usepackage{graphicx}
\usepackage{latexsym}
\usepackage{amsmath}
\usepackage{amssymb}
\usepackage{multirow}
\usepackage{bm}
\usepackage{url}

\setcitestyle{square,numbers,comma,sort&compress}

\newcommand{\rpm}{\raisebox{.2ex}{$\scriptstyle\pm$}}

\begin{document}

\title{Attention U-Net Based Adversarial Architectures for Chest X-ray Lung Segmentation}

\author{
Guszt\'av Ga\'al \hskip6.5em Bal\'azs Maga \hskip6.5em Andr\'as Luk\'acs\\
{\small \texttt{guzzzti@gmail.com}} \hskip2em {\small \texttt{mbalazs0701@gmail.com}} \hskip2em {\small \texttt{lukacs@cs.elte.hu}}\\
AI Research Group, Institute of Mathematics\\
E\"otv\"os Lor\'and University, Budapest, Hungary
}

\maketitle
\bibliographystyle{unsrt}

\begin{abstract}
Chest X-ray (CXR) is the most common test among medical imaging modalities. It is applied for detection and differentiation of, among others, lung cancer, tuberculosis, and pneumonia, the last with importance due to the COVID-19 disease. Integrating computer-aided detection methods into the radiologist diagnostic pipeline, greatly reduces the doctors' workload, increasing reliability and quantitative analysis. Here we present a novel deep learning approach for lung segmentation, a basic, but arduous task in the diagnostic pipeline. Our method uses state-of-the-art fully convolutional neural networks in conjunction with an adversarial critic model. It generalized well to CXR images of unseen datasets with different patient profiles, achieving a final DSC of 97.5\% on the \textit{JSRT} dataset.
\end{abstract}

\section{Introduction}

X-ray is the most commonly performed radiographic examination, being significantly easier to access, cheaper and faster to carry out than computed tomography (CT), diagnostic ultrasound and magnetic resonance imaging (MRI), as well as having lower dose of radiation compared to a CT scan. According to the publicly available, official data of the National Health Service (\cite{england2019diagnostic}), in the period from February 2017 to February 2018, the count of imaging activity was about 41 million in England, out of which almost 22 million was plain X-ray. Many of these imaging tests might contribute to early diagnosis of cancer, amongst which chest X-ray is the most commonly requested one by general practitioners. In order to identify lung nodules, lung segmentation of chest X-rays is essential, and this step is vital in other diagnostic pipelines as well, such as calculating the cardiothoracic ratio, which is the primary indicator of cardiomegaly. For this reason, a robust algorithm to perform this otherwise arduous segmentation task is much desired in the field of medical imaging.

Semantic segmentation aims to solve the challenging problem of assigning a pre-defined class to each pixel of the image. This task requires a high level of visual understanding, in which state-of-the-art performance is attained by methods utilizing Fully Convolutional Networks (FCN) \cite{long2015fully}. In \cite{luc2016semantic}, adversarial training is used to enhance segmentation of colored images. This idea was incorporated to \cite{wei2018scan} in order to segment chest X-rays with a fully convolutional, residual neural network. Recently, Mask R-CNN \cite{he2017mask} is utilized to realize instance segmentation on chest X-rays and obtained state-of-the-art results \cite{wang2019instance,hu2020effective}.

\section{Deep Learning Approach}

\subsection{Network Architecture}

\begin{figure}[h!]
\centering 
\includegraphics[scale=0.6]{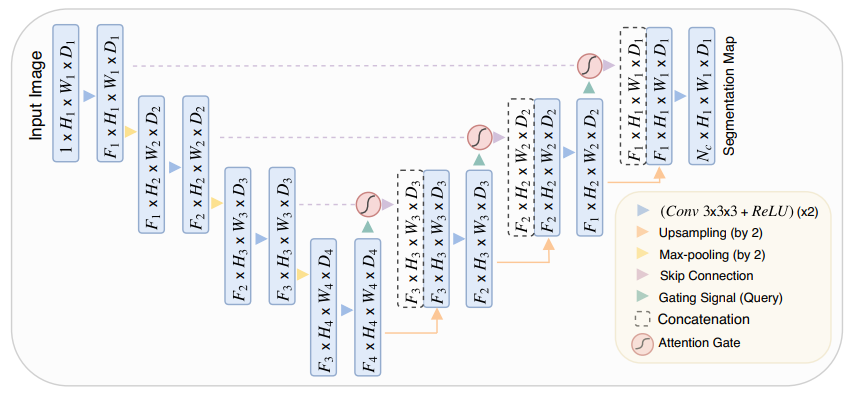}
\caption{Schematic architecture of the Attention U-Net \cite{oktay2018attention}}
\label{attention_u-net}
\end{figure}

Our goal is to produce accurate organ segmentation masks on chest X-rays, meaning for input images we want pixel-wise dense predictions regarding if the given pixel is either part of the left lung, the right lung, the heart, or none of the above. \newline
For this purpose Fully Convolutional Networks (FCNs) are known to significantly outperform other widely used registration-based methods. Specifically we applied a U-Net architecture, thus enabling us to efficiently compute the segmentation mask in the same resolution as the input images. The fully convolutional architecture also enables the use images of different resolutions, since unlike standard convolutional networks, FCNs don't contain input-size dependent layers. \newline
In \cite{oktay2018attention} it has been shown that for medical image analysis tasks the integration of the proposed Attention Gates (AGs) improved the accuracy of the segmentation models, while preserving computational efficiency. The architecture of the proposed Attention U-Net is described by Figure \ref{attention_u-net}. Without the use of AGs, it's common practice to use cascade CNNs, selecting a Region Of Interest (ROI) with another CNN where the target organ is likely contained. With the use of AGs we eliminate the need for such a preselecting network, instead the Attention U-Net learns to focus on most important local features, and dulls down the less relevant ones. We note that the dulling of less relevant local features also result in decreased false positive rates.

In order to enhance the performance of Attention U-Net, we further experimented with adversarial techniques, motivated by \cite{wei2018scan}. In that work, the authors first designed a Fully Convolutional Network (FCN) for the lung segmentation task, and noted that in certain cases the network tends to segment abnormal and incorrect organ shapes. For example, the apex of the ribcage might be mistaken as an internal rib bone,  resulting in the mask “bleeding out” to the background, which has similar intensity as the lung field. To address this issue, they developed an adversarial scheme, leading to a model which they call Structure Correcting Adversarial Network (SCAN).
\begin{figure}[h!]
\centering 
\includegraphics[scale=0.4]{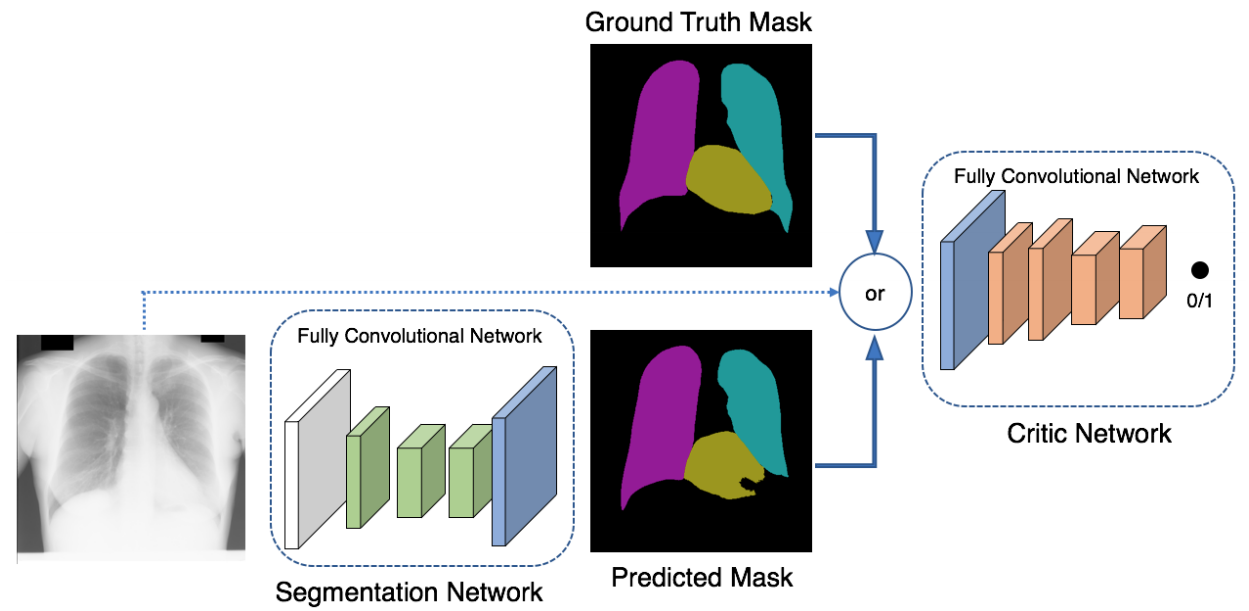}
\caption{Schematic architecture of the Structure Correcting Adversarial Networks \cite{wei2018scan}}
\label{scan_structure}
\end{figure}
This architecture is based on the idea of the General Adversarial Networks \cite{goodfellow2014generative}. They use the pretrained Fully Convolutional Network as a generator of a General Adversarial Network, and they also train a critic network which is fed the ground truth mask, the predicted mask and optionally the original image. The critic network has roughly the same architecture, resulting in similar capacity. This approach forces the generator to segment more realistic masks, eventually removing obviously wrong shapes.

In our work, besides the standard Attention U-Net, we also created a network of analogous structure, in which the FCN used in \cite{wei2018scan} is replaced by the Attention U-Net. We did not introduce any modification in the critic model design, such experiments are left to future work.

\subsection{Tversky Loss}

In the field of medical imaging, Dice Score Coefficient (DSC) is probably the most widespread and simple way to measure the overlap ratio of the masks and the ground truth, and hence to compare and evaluate segmentations. Given two sets of pixels $X,Y$, their DSC is
\begin{displaymath}
DSC(X,Y) = \frac{ 2|X\cap Y|}{|X|+|Y|}.
\end{displaymath}
If $Y$ is in fact the result of a test about which pixels are in $X$, we can rewrite it with the usual notation true/false positive (TP/FP), false negative (FN) to be
\begin{displaymath}
DSC(X,Y) = \frac{2TP}{2TP + FN + FP}.
\end{displaymath}
We would like to use this concept in our setup. The class $c$ we would like to segment corresponds to a set, but it is more appropriate to consider its indicator function $g$, that is $g_{i,c}\in \{0,1\}$ equals 1 if and only if the $i$th pixel belongs to the object. On the other hand, our prediction is a probability for each pixel denoted by $p_{i,c}\in[0,1]$. Then the Dice Score of the prediction in the spirit of the above description is defined to be
\begin{displaymath}
DSC = \frac{\displaystyle\sum_{i=1}^N p_{i,c}g_{i,c} + \varepsilon}{\displaystyle\sum_{i=1}^N \left(p_{i,c}+g_{i,c}\right) + \varepsilon},
\end{displaymath}
where $N$ is the total number of pixels, and $\varepsilon$ is introduced for the sake of numerical stability and to avoid divison by 0. The linear Dice Loss (DL) of the multiclass prediction is then
\begin{displaymath}
DL = \sum_{c}\left(1 - DSC_c\right).
\end{displaymath}
A deficiency of Dice Loss is that it penalizes false negative and false positive predictions equally, which results in high precision but low recall. For example practice shows that if the region of interests (ROI) are small, false negative pixels need to have a higher weight than false positive ones. Mathematically this obstacle is easily overcome by introducing weights $\alpha,\beta$ as tuneable parameters, resulting in the definition of Tversky similarity index \cite{tversky1977features}:
\begin{displaymath}
TI_c = \frac{\displaystyle\sum_{i=1}^N p_{i,c}g_{i,c} + \varepsilon}{\displaystyle\sum_{i=1}^N p_{i,c}g_{i,c} + \alpha\displaystyle\sum_{i=1}^N p_{i,\overline{c}}g_{i,c}+\beta\displaystyle\sum_{i=1}^N p_{i,c}g_{i,\overline{c}} + \varepsilon},
\end{displaymath}
where $p_{i,\overline{c}}=1-p_{i,c}$ and $g_{i,\overline{c}}=1-g_{i,c}$, that is the overline simply stands for describing the complement of the class. \\
Tversky Loss is obtained from Tversky index as Dice Loss was obtained from Dice Score Coefficient:
\begin{displaymath}
TL = \sum_{c}\left(1 - TI_c\right).
\end{displaymath}
Another issue with the Dice Loss is that it struggles to segment small ROIs as they do not contribute to the loss significantly. This difficulty was addressed in \cite{abraham2019novel}, where the authors introduced the quantity Focal Tversky Loss in order to improve the performance of their lesion segmentation model:
\begin{displaymath}
FTL = \sum_{c}\left(1 - TI_c\right)^{\gamma^{-1}},
\end{displaymath}
where $\gamma \in [1,3]$. In practice, if a pixel with is misclassified with a high Tversky index, the Focal Tversky Loss is unaffected. However, if the Tversky index is small and the pixel is misclassified, the Focal Tversky Loss will decrease significantly. 

\subsection{Training}

The explanation of the training of our structure correcting network is a bit longer to explain, we directly follow the footsteps of \cite{wei2018scan}. Let $S$, $D$ be the segmentation network and the critic network, respectively. The data consist of the input images $\bm{x}_i$ and the associated mask labels $\bm{y}_i$, where $\bm{x}_i$ is of shape $[H, W, 1]$ for a single-channel gray-scale image with height $H$ and width $W$, and $\bm{y}_i$ is of shape $[H,W,C]$ where $C$ is the number of classes including the background. Note that for each pixel location $(j,k)$, $y_i^{jkc}=1$ for the labeled class channel $c$ while the rest of the channels are zero ($y_i^{jkc'}=0$ for $c'\ne c$). We use $S(\bm{x}) \in [0,1]^{[H,W,C]}$ to denote the class probabilities predicted by $S$ at each pixel location such that the class probabilities normalize to 1 at each pixel. Let $D(\bm{x}_i, \bm{y})$ be the scalar probability estimate of $\bm{y}$ coming from the training data. They defined the optimization problem as
\begin{equation}
\min_S \max_D \Bigl\{\sum_{i=1}^N J_s(S(\bm{x}_i), \bm{y}_i) - \lambda \Big[ J_d(D(\bm{x}_i, \bm{y}_i), 1) + J_d(D(\bm{x}_i, S(\bm{x}_i)), 0) \Big]  \Bigr\}, 
\label{eq:objective}
\end{equation}
 where $$J_s(\bm{\hat{y}},\bm{y}) := \frac{1}{HW}\sum_{j,k}\sum_{c=1}^C -y^{jkc}\ln y^{jkc}$$ is the multiclass cross-entropy loss for predicted mask $\bm{\hat{y}}$ averaged over all pixels. $$J_d(\hat{t}, t) := -t \ln \hat{t} + (1-t)\ln(1-\hat{t})$$ is the binary logistic loss for the critic's prediction. $\lambda$ is a tuning parameter balancing pixel-wise loss and the adversarial loss. We can solve equation~\eqref{eq:objective} by alternate between optimizing $S$ and optimizing $D$ using their respective loss functions. This is a point where we introduced a modification: instead of using the multiclass cross-entropy loss $J_s(\bm{\hat{y}},\bm{y})$ in the first term, we applied the Focal Tversky Loss $FTL(\bm{\hat{y}},\bm{y})$.

Now since the first term in equation~\eqref{eq:objective} does not depend on $D$, we can train our critic network by {\it minimizing} the following objective with respect to $D$ for a fixed $S$:
\begin{equation*}
\footnotesize
\sum_{i=1}^N J_d(D(\bm{x}_i, \bm{y}_i), 1) + J_d(D(\bm{x}_i, S(\bm{x}_i)), 0).
\end{equation*}
Moreover, given a fixed $D$, we train the segmentation network by minimizing the following objective with respect to $S$:
\begin{equation*}
\footnotesize
\sum_{i=1}^N FTL(S(\bm{x}_i), \bm{y}_i) + \lambda J_d(D(\bm{x}_i, S(\bm{x}_i)), 1).
\end{equation*}
Following the recommendation in~\cite{goodfellow2014generative}, we use $J_d(D(\bm{x}_i, S(\bm{x}_i)), 1)$ in place of $-J_d(D(\bm{x}_i, S(\bm{x})), 0)$,  as it leads to stronger gradient signals. After tests on the value of $\lambda$ we decided to use $\lambda=0.1$.

Concerning training schedule, we found that following pretraining the generator for 50 epochs, we can train the adversarial network for 50 epochs, in which we perform 1 optimization step on the critic network after each 5 optimization step on the generator. This choice of balance is also borrowed from \cite{wei2018scan}, however, we note that the training of our network is much faster.

\section{Datasets}
For training- and validation data, we used the Japanese Society of Radiological Technology (JSRT) dataset \cite{shiraishi2000development} , as well as the Montgomery- and Shenzhen dataset \cite{jaeger2014two}, all of which are public datasets of chest X-rays with available organ segmentation masks reviewed by expert radiologists. \newline
The JSRT dataset contains a total of 247 images, of which 154 contains lung nodules. The X-rays are all in $2048 \times 2048$ resolution, and have 12-bit grayscale levels. Both lung and heart segmentation masks are available for this dataset. \newline
The Montgomery dataset contains 138 chest X-rays, of which 80 X-rays are from healthy patients, and 58 are from patients with tuberculosis. The X-rays have either a resolution of $4020 \times 4892$ or $4892 \times 4020$, and have 12-bit grayscale levels as well. In the case of this dataset, only lung segmentation masks are publicly available. \newline
The Shenzhen dataset contains a total of 662 chest X-rays, of which 326 are of healthy patients, and in a similar fashion, 336 are of patients with tuberculosis. The images vary in sizes, but all are of high resolution, with 8-bit grayscale levels. Only lung segmentation masks are publicly available for the dataset.

\subsection{Preprocessing Data}

\begin{figure}[h!]
\centering 
\includegraphics[scale=0.5]{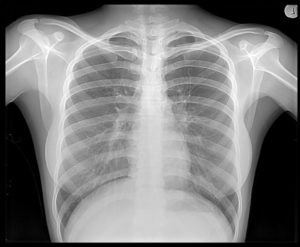}
\includegraphics[scale=0.5]{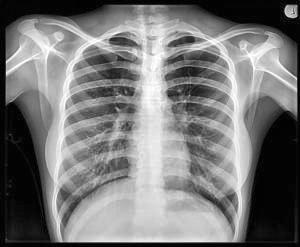}
\caption{Example of chest X-ray images before and after CLAHE}
\label{claheimg}
\end{figure}

X-rays are grayscale images with typically low contrast, which makes their analysis a difficult task. This obstacle might be overcome by using some sort of histogram equalization technique. The idea of standard histogram equalization is spreading out the the most frequent intensity values to a higher range of the intensity domain $[0, 255]$ by modifying the intensities so that their cumulative distribution function (CDF) on the complete modified image is as close to the CDF of the uniform distribution as possible. Improvements might be made by using adaptive histogram equalization, in which the above method is not utilized globally, but separately on pieces of the image, in order to enhance local contrasts. However, this technique might overamplify noise in near-constant regions, hence our choice was to use Contrast Limited Adaptive Histogram Equalization (CLAHE), which counteracts this effect by clipping the histogram at a predefined value before calculating the CDF, and redistribute this part of the image equally among all the histogram bins. \\
Applying CLAHE to an X-ray image has visually appealing results, as displayed in Figure \ref{claheimg}. As our experiments displayed, it does not merely help human vision, but also neural networks.

The images were then resized to 512x512 resolution and mapped to $[-1,1]$ before being fed to our network.

\section{Experiments and Results}

The aforementioned Attention U-Net architecture was implemented using Keras and TensorFlow Python neural-network libraries, to which we have fed our dataset and trained for $40$ epochs with $8$ X-ray scans in each batch. Our optimizer of choice was Stochastic Gradient Descent, having found that Adam failed to converge in many cases. As loss function, we applied Focal Tversky Loss.

We have found that applying various data augmentation techniques such as flipping, rotating, shearing the image as well as increasing or decreasing the brightness of the image were of no help and just resulted in slower convergence.

Using the Attention U-Net infrastructure, we managed to reach a dice score of 0.962 for the segmentation of the lung. Unlike in \cite{wei2018scan}, where no major preprocessing was done, with our preprocessing method, the network performed very well even if the test- and the validation sets were of different datasets. This is extremely important for real world applications, as X-ray images of different machines are significantly different, largely dependent on the specific calibration of each machine, thus it is no trivial task to have X-rays accurately evaluated that are from machines from which no images were in the training set. \newline

\begin{table}[h!]
\begin{center}
{\caption{Dice scores of different architectures over different datasets.}\label{table1}}
\begin{tabular}{lccc}
\hline
\rule{0pt}{12pt}
Dataset&$\textbf{SCAN}$&$\textbf{ATTN U-Net}$&$\textbf{Ours (Adv. ATTN)}$
\\
\hline
\\[-6pt]
\quad\textbf{JSRT}&97.3 \rpm 0.8\%&96.3 \rpm 0.7\%&\textbf{97.6 \rpm 0.5\%}\\
\quad\textbf{All}&-&95.8 \rpm 0.4\%&\textbf{96.2 \rpm 0.4\%}\\
\quad\textbf{All /\ JSRT}&-&96.6 \rpm 0.6&\textbf{97.8 \rpm 0.6\%}
\\
\end{tabular}
\end{center}
\end{table}

\begin{figure}[h!]
\centering 
\includegraphics[scale=0.6]{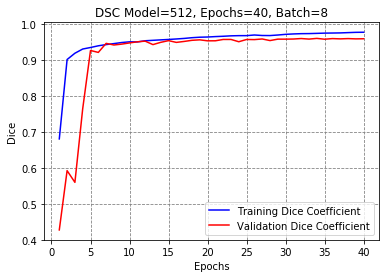}
\caption{Epoch-wise dice score coefficient}
\label{lossplot}
\end{figure}

We note that even though introducing the adversarial scheme in our setting increased the dice scores, the improvement was not as drastic as in the case of the FCN and SCAN. By checking the masks generated by the vanilla Attention U-Net, we found that this phenomenon can be attributed to the fact that while the FCN occasionally produces abnormally shaped masks, due to our preprocessing steps the Attention U-Net does not commit this mistake. Consequently, the adversarial scheme is responsible for subtle shape improvements only, which is indicated by the Dice Score less spectacularly.


\section{Future Work}

So far we have not experimented with the architecture of the critic network, we found the performance of the architecture in \cite{wei2018scan} completely satisfying. However, it would be desirable to carry out further tests in this direction in order to achieve better understanding of the role of adversarial scheme.

\ack The project was partially supported by the AI4EU project, funded by EU H2020 programme (contract no. 825619). Furthermore, it was supported by the Hungarian National Excellence Grant 2018-1.2.1-NKP-00008 and the grant EFOP-3.6.3-VEKOP-16-2017-00002. The last author was supported by project no. ED\_18-1-2019-0030 (Application domain specific highly reliable IT solutions) funded from the NRDI Fund of Hungary, under the Thematic Excellence Programme scheme.

\bibliography{attention-unet}
\end{document}